\documentstyle[aps,prl,preprint,epsf,rotate]{revtex}
\tightenlines
\begin{document}
\title{Quasiparticle resonant states as a probe of short-range
electronic structure and Andre\'ev coherence}
\author{Michael E. Flatt\'e}
\address{Department of Physics and Astronomy, University of Iowa, 
Iowa City, IA 52242}
\maketitle
\begin{abstract}
The recently observed properties of quasiparticle resonant states 
near impurities on the surface of superconducting 
Bi$_2$Sr$_2$CaCu$_2$O$_{8+\delta}$ demonstrate that in-plane Andre\'ev
processes are either absent or phase-incoherent. Analysis of
the spectral and 
spatial details of the electronic structure near a Zn impurity also
suggest an effective magnetic component of the impurity
potential.  Further 
experiments are proposed to clarify whether the effective
moments of nearby impurities are correlated.
\end{abstract}
\vfill\eject

Over the past few years several authors have emphasized the wealth 
of information available from local probes of impurity properties in 
correlated electron systems, and particularly in superconductors 
whose homogeneous order parameters (OP) are 
anisotropic in momentum\cite{FBSSP}. 
A parallel improvement in scanning tunneling spectroscopy
(STS) has allowed 
this vision to become a reality through direct observation
of the local density of states (LDOS), 
first in niobium\cite{YazdaniNb}, which has a 
momentum-independent OP, and this 
year in the high-temperature 
superconductor Bi$_2$Sr$_2$CaCu$_2$O$_{8+\delta}$
 (BSCCO)\cite{Hudson,YazdaniBSC,Pan}, which has 
an anisotropic OP. 
The electronic structure of the BSCCO surface
is much more complex than that of the 
niobium surface; there are local moments in the copper-oxygen 
planes and, under certain conditions, a pseudogap state.
Recent work\cite{pseudogap}
has emphasized the role of the 
pseudogap state\cite{bulk,phaseinc,stripe} 
in determining the properties of clean surfaces of
high-temperature superconductors at temperatures near T$_c$, and STS and
photoemission have directly demonstrated its existence on
the surface of BSCCO. The pseudogap state
is characterized by a single-particle gap of $d_{x^2-y^2}$
symmetry, but the Andre\'ev processes one would expect in
a superconducting state are (according to differences in
mechanism) either absent or
phase-incoherent. 

Here the STS measurements near
impurities\cite{Hudson,YazdaniBSC,Pan} will be shown to 
unambiguously demonstrate essentially complete
suppression of Andre\'ev processes in the vicinity of the
impurity, even at very low temperatures.  
This low-temperature indication of a pseudogap implies its
importance to the nature of
the superconducting state and the operation of devices (such
as Josephson junctions) with such materials. Indications of an
effective magnetic component can also be seen in Ref.~\cite{Pan}. 
Thus STS provides a direct probe of both the
local Andre\'ev (superconducting) coherence and the local magnetic
properties on the BSCCO surface.

A brief review of the experimental results from
Refs.~\cite{Hudson,YazdaniBSC,Pan} is in order.
Theoretical predictions which have been confirmed include
the presence of quasiparticle resonances near 
nonmagnetic impurities in anisotropic OP 
superconductors\cite{SBS}, as well as the
suppression of the gap feature near the 
impurity and the
asymmetry of the resonance peak due to the energy dependence of the 
quasiparticle density of states\cite{FB2}. The 
disagreements with previous theory, however, are striking. 
The most noticeable one
is that the resonant state has only been detected on the {\it
hole } side of the spectrum, both on the impurity site and {\it 
everywhere else around the impurity.}
Whereas previous theories are consistent with an LDOS
measured at the impurity which is entirely hole-like, these same theories 
unambiguously predict the LDOS at nearest-neighbor states will be almost 
entirely electron-like.  Indeed these theories predict the 
spatially-integrated LDOS (or DOS) will be nearly
particle-hole symmetric even 
though the LDOS at any particular site is not.
A second unexpected element in the data is
the presence of a second, much smaller,
spectral peak on the hole side in Ref.~\cite{Pan}. 
Whereas 
previous theories are consistent with two resonances at a single impurity,
the spatially-integrated amplitude of each resonance should be 
approximately the same, unlike what is seen in Ref.~\cite{Pan}. 

The above two issues 
are disagreements between the 
experimental and theoretical DOS, but there are also two significant 
disagreements in the LDOS. The resonance has a large amplitude at the 
impurity site, whereas calculations indicate that the largest amplitude in 
the LDOS should occur at the nearest-neighbor sites. Finally,
the gap
feature is seen on the impurity site, where it does not
appear in calculations.\cite{FBSSP}
 
The calculations here incorporate the pseudogap state
into the evaluation of the differential conductance
($dI/dV$). 
After evaluating the LDOS for several
impurity potential models, allowing for spatial extent and
magnetic character in the potential,
the impurities 
of Refs~\cite{Hudson,YazdaniBSC,Pan} are found to be highly localized
and the four discrepancies above can be reconciled.
Finally, in the case of Ref.~\cite{Pan}, 
a magnetic component to the effective Zn impurity potential is
evident.\cite{Znmag}

The calculations of the LDOS are based on the
Hamiltonian
\begin{equation}
H = \sum_{\langle ij\rangle,\sigma}
\left[-t_{ij}c_{i\sigma}^\dagger c_{
j\sigma} + \Delta_{ij}
c_{i\uparrow}^\dagger c_{j\downarrow}^\dagger+\Delta_{ij}^*
c_{j\downarrow}c_{i\uparrow}\right] + \sum_i\left[
(V_{0i}+V_{Si})c_{i\uparrow}^\dagger
c_{i\uparrow} + (V_{0i}-V_{Si})
c_{i\downarrow}^\dagger c_{i\downarrow}\right],
\end{equation}
which
includes a site-dependent potential which can be magnetic
($V_S$), nonmagnetic ($V_0$), or a combination of both.  $i$
and $j$ label sites and $\sigma$ labels spin. 
The homogeneous electronic structure, expressed as hopping
matrix elements ($t_{ij}$) for the first five nearest neighbors,
is taken from a
single-band parametrization of photoemission data.\cite{BSCnormal}
Large variations in hopping matrix elements ($>50$~meV)
produce results very much at odds with experiment, whereas
smaller variations are mimicked by the site-dependent
potential. Hence such changes will be ignored here.
Only on-site and nearest-neighbor order parameters $\Delta_{ij}$ are
nonzero, and the maximum OP on the Fermi surface, 
$\Delta_{max} = 40$~meV.\cite{OPmax}

\begin{figure}
\epsfxsize=5.0in
\rotate[r]{\epsffile{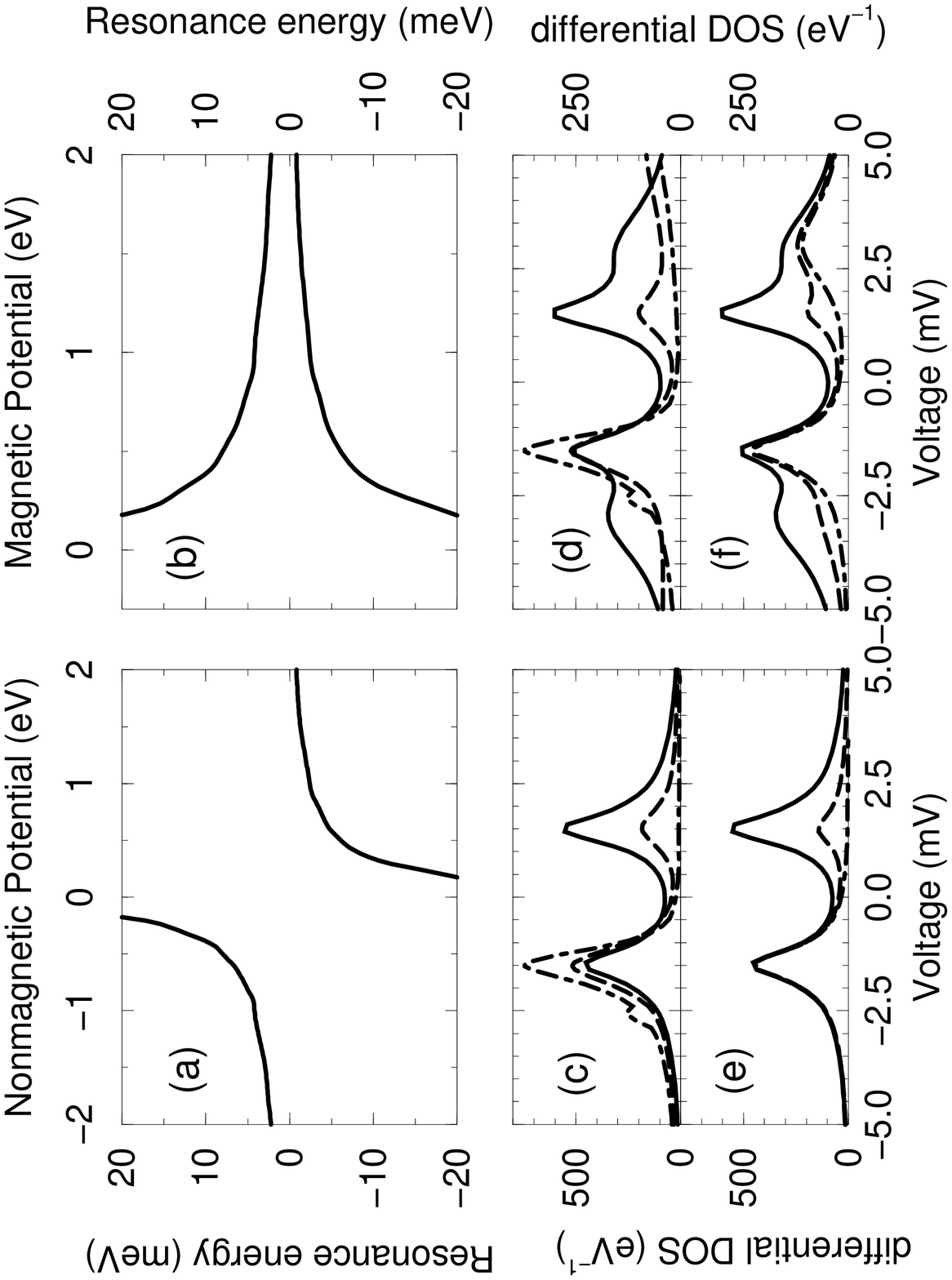}}
\caption[]{Resonance energies for single-site (a) nonmagnetic and
(b) magnetic impurity potentials. DOS for (c) nonmagnetic
and (d) magnetic impurity potentials which produce a
resonance at -1.5mV. Solid, $40$~meV superconducting gap, 
dashed, $25$~meV superconducting, $40$~meV total gap, 
dot-dashed, $40$~meV non-superconducting gap. (e) and (f), same as (c)
and (d) except the dashed line corresponds to partial and the dot-dashed
line to no phase coherence in the $40$~meV superconducting 
system.  \label{resonance}}
\end{figure}

The electronic 
structure of the inhomogeneous system (including the impurity) is 
determined by direct numerical solution in real space of the Gor'kov 
equation (in Nambu form) for the inhomogeneous Green's function,
${\bf G} = {\bf g} + {\bf
g}{\bf V}{\bf G} = ({\bf I}-{\bf g}{\bf V})^{-1}{\bf g}$,
within a 
real-space region around the impurity 
beyond which the potential is negligible. \cite{FBSSP}
The $\Delta_{ij}$'s are found self-consistently
in this process, for they
determine the off-diagonal components of the potential ${\bf
V}$.  Spectra outside this real-space
region 
are constructed according
to the generalized ${\bf T}$-matrix equation: 
${\bf G} = {\bf g} + {\bf g}{\bf V}
\left[{\bf I - GV}\right]^{-1}{\bf g}$. 
Once ${\bf G}$ has been calculated throughout the region
near the impurity, the LDOS and DOS are obtained from its
imaginary part and the lattice Wannier functions. Then
\begin{equation}
{dI({\bf x};V)\over dV} = -\int d\omega\sum_\sigma {1\over \pi}
\left({\partial n_{STM}(\omega)\over \partial \omega}\right)
|\phi_\sigma({\bf
x};i)|^2{\rm Im}G_\sigma(i,i;\omega),
\end{equation}
where $\phi_\sigma({\bf x};i)$ is the overlap of the Wannier
function at site $i$ and spin $\sigma$ with the STM tip at
${\bf x}$, and $n_{STM}(\omega)$ is the occupation function
of the STM tip.\cite{FBSSP}
Resonances correspond to new peaks in the differential DOS
(the difference between the inhomogeneous and homogeneous
DOS); their
energies are shown in Fig.~\ref{resonance}(ab) for magnetic and
nonmagnetic single-site impurities. 

If Andre\'ev processes are suppressed, either by reduction
in their amplitude or phase coherence, a resonance's DOS
will become more electron-like or more hole-like.
Reduction of the amplitude of the homogeneous anomalous Green's
function $f(i,j;\omega)$, due
perhaps to a local antiferromagnetic (AF) order,
decreases the mean-field coupling between electron and hole
excitations. 
Note that this is very different from the fully electron or
hole-like character of the {\it LDOS} at the impurity, which
originates from a vanishing $f(i, i;\omega)$ in
the $d_{x^2-y^2}$ state. Figure~\ref{resonance}(cd) shows the DOS of a
resonance for three systems with a 40~meV $d_{x^2-y^2}$ gap: a fully
superconducting gap (solid line), a gap with a 25~meV
superconducting component (dashed line), and a pseudogap with
no superconducting component (dot-dashed line). As the
superconducting component is reduced, the electron-hole
symmetry diminishes. The nonmagnetic potentials of Fig.~\ref{resonance}(c)
are chosen ($1.375$~eV, $1.000$~eV, and $0.833$~eV, respectively) so
the resonance peak is at $-1.5$~mV (the same as
Ref.~\cite{Pan}). The magnetic potentials in Fig.~\ref{resonance}(d) are
the same as those in Fig.~\ref{resonance}(c).

The reduction of the electron-like peak from
phase decoherence is similar to the 
effect of amplitude suppression of $f(i,i;\omega)$. 
For a resonance at $\omega$, the peak at $-\omega$ comes from
terms with products of pairs of anomalous Green's functions.
For a phase incoherent pseudogap state the expectation value
of these pairs, and thus the amplitude of the electron-like
peak, is diminished.\cite{phaseinc} The effect of this is shown in
Fig.~\ref{resonance}(ef) as a dashed line corresponding
to partial (half) and a dot-dashed line corresponding to no phase coherence.
The nonmagnetic potential is $1.375$~eV in
Fig.~\ref{resonance}(e) and the magnetic potential 
is $1.375$~eV in Fig.~\ref{resonance}(f).
Note that for the purely magnetic impurities even in the
absence of Andre\'ev coherence there is a peak on each 
side of zero energy.
 
For Refs.~\cite{Hudson,Pan} there is no apparent
electron-like component of the resonance in the DOS,
thus local Andre\'ev coherence is absent.
In measurements of the LDOS near metal islands on BSCCO\cite{YazdaniBSC}, 
however, both hole-like and electron-like peaks
are apparent. This may indicate that the metal plays an important role 
in maintaining phase 
coherence at the surface, or that the metal overlayer is less 
disruptive to superconductivity 
than impurities in the plane. The presence of the in-plane
Andre\'ev processes, indicated by the electron-like peak, is
essential to the operation of Josephson junctions.

\begin{figure}
\epsfxsize=5.0in
\rotate[r]{\epsffile{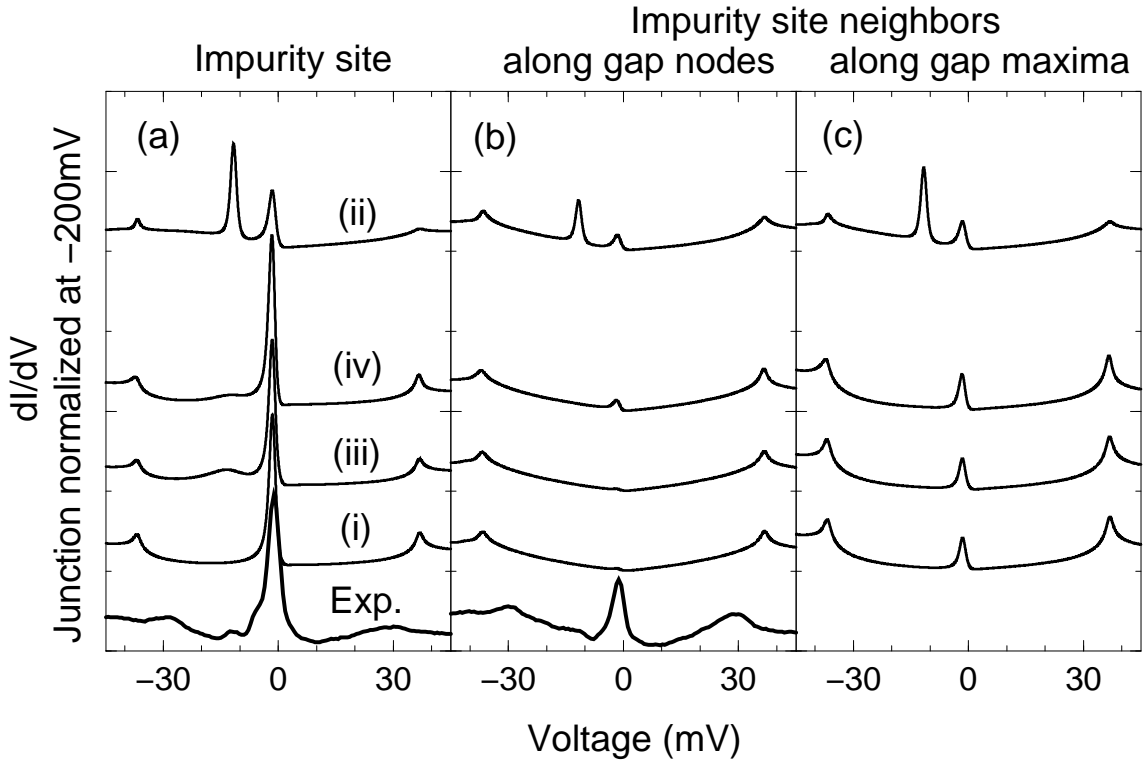}}
\caption[]{$dI/dV$  (a) on-site, (b) at the nearest-neighbor along the
gap nodes, and (c) along the gap maxima,
for impurity and host models {\it i}, {\it ii},
{\it iii}, and {\it iv}. Data of
Ref.~\cite{Pan} is also shown.  The order of vertical offsets (introduced
for clarity) of
the curves in (b) and (c) is the same as in (a).\label{magspace}}
\end{figure}

The on-site LDOS of Ref.~\cite{Pan} is shown in Fig.~\ref{magspace}(a).
The second (hole-like) peak is not as clearly 
evident in the results of Refs.~\cite{Hudson,YazdaniBSC}, 
and thus may be peculiar either to the Zn impurity or to the impurity site in 
the BSCCO unit cell. Additional resonances around 
impurities can originate from additional orbital states around
spatially-extended potentials or from spin-splitting near
magnetic potentials. 
Note that an effective magnetic potential could also originate from a 
nonmagnetic impurity potential placed in a spin-polarized host electronic 
structure. 

Figure~\ref{magspace}(abc) shows 
the  best fit of $dI/dV$
to the data of Ref.~\cite{Pan} for phase
incoherent Andre\'ev processes and {\it i} a
single-site nonmagnetic potential (1.375~eV), {\it ii}
a nonmagnetic potential with onsite (0.360~eV) and
nearest-neighbor (0.150~eV) values, and {\it iii} 
a mixed nonmagnetic and
magnetic potential ($V_0 = 0.825$~eV, $V_S = 0.550$~eV). Also
shown is the best fit using {\it iv} a pseudogap with no
superconducting component
and a mixed potential ($V_0 = 0.543$~eV, $V_S = 0.290$~eV).
The three panels show $dI/dV$ (a) at the impurity site, (b)
at the nearest-neighbor site along the gap nodes,
and (c) along the gap maxima.
Measurements of Ref.~\cite{Pan} for (a) and (b) are shown;
(c) is not available.

The large size of the resonance on-site and the simultaneous
presence of the gap feature occur because of junction
normalization (equal resistance at $-200$~mV) 
and the finite width of the $|\phi_\sigma|^2$ (modeled as Gaussians
of range $0.8$\AA, $3.8$\AA, $0.8$\AA, and $1.0$\AA\ for
{\it i-iv}).
The very small LDOS in this energy range
at the impurity site causes the tip to approach closer to
the surface and (1) enlarge the apparent size of the resonance on-site, and
(2) pick up the gap features from the nearest-neighbor sites. 
The relative
size of the on-site resonance to the gap features is largely determined by the
overlap of the nearest-neighbor Wannier functions with the tip when the tip is
over the impurity site.

Judging from the comparison with experiment, 
{\it iii} and {\it iv} appear most in agreement.
{\it i} does not have a second resonance,
and whereas {\it ii} does show one
in the proper location, its relative magnitude is incorrect.
The smaller amplitude of the second resonance is obtained for 
{\it iii} and {\it iv}
because the overlaps with the STM tip are spin-dependent 
($|\phi_\uparrow({\bf x})|^2/|\phi_\downarrow({\bf x})|^2\sim 40$).
If the second peak were absent, either the impurity would
lack magnetic character, or it would
occur in less magnetic regions of the BSCCO unit cell. 
The remaining disagreement is in the amplitude of the
resonance in (B), where {\it iv} is best, but still too
small. 

\begin{figure}
\epsfxsize=5.0in
\rotate[r]{\epsffile{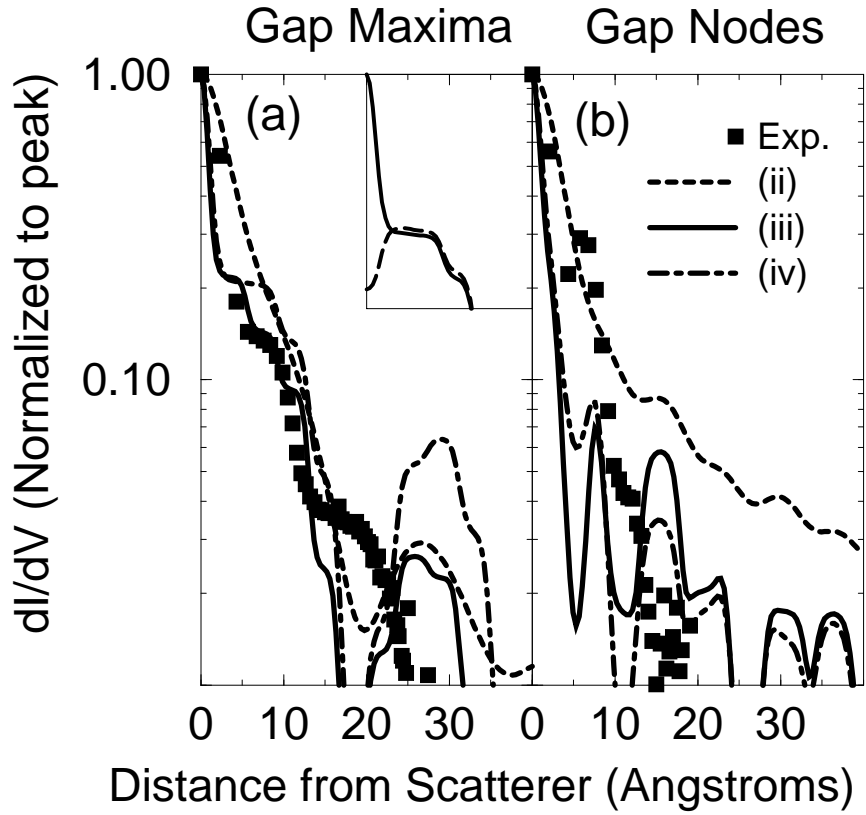}}
\caption[]{$dI/dV$ as a function of position along (a)
the gap maxima
and (b) the gap nodes. {\it ii} is dotted, {\it iii} is solid,
and {\it iv} is dot-dashed.
Shown in inset is the difference between $dI/dV$ when junction
normalization is (solid) and is not (dashed)
taken into account.
\label{lineslices} }
\end{figure}

Figure~\ref{lineslices} shows the amplitude of the resonance
as a function of distance from the impurity 
along the gap maxima (a) and the gap nodes (b). The 
squares are the data from Ref.~\cite{Pan}, whereas the solid line 
corresponds to {\it iii}, the dotted line to {\it ii},
and the dot-dashed line to {\it iv}. The plot for {\it i} looks
identical to that of {\it iii}.
The agreement of {\it iii} and {\it iv} are quite good along
the maxima direction. The absence of a well-defined maximum
at the nearest-neighbor in (a), which was pointed out in
Ref.~\cite{Pan}, is due to the normalization procedure. 
An inset in Fig.~\ref{lineslices} shows the difference between the 
junction normalized 
(solid) and unnormalized (dashed) $dI/dV$. The main discrepancy
is with the amplitude of the signal along the node
directions (b). 

\begin{figure}
\epsfxsize=10.0in
{\epsffile[50 400 662 792]{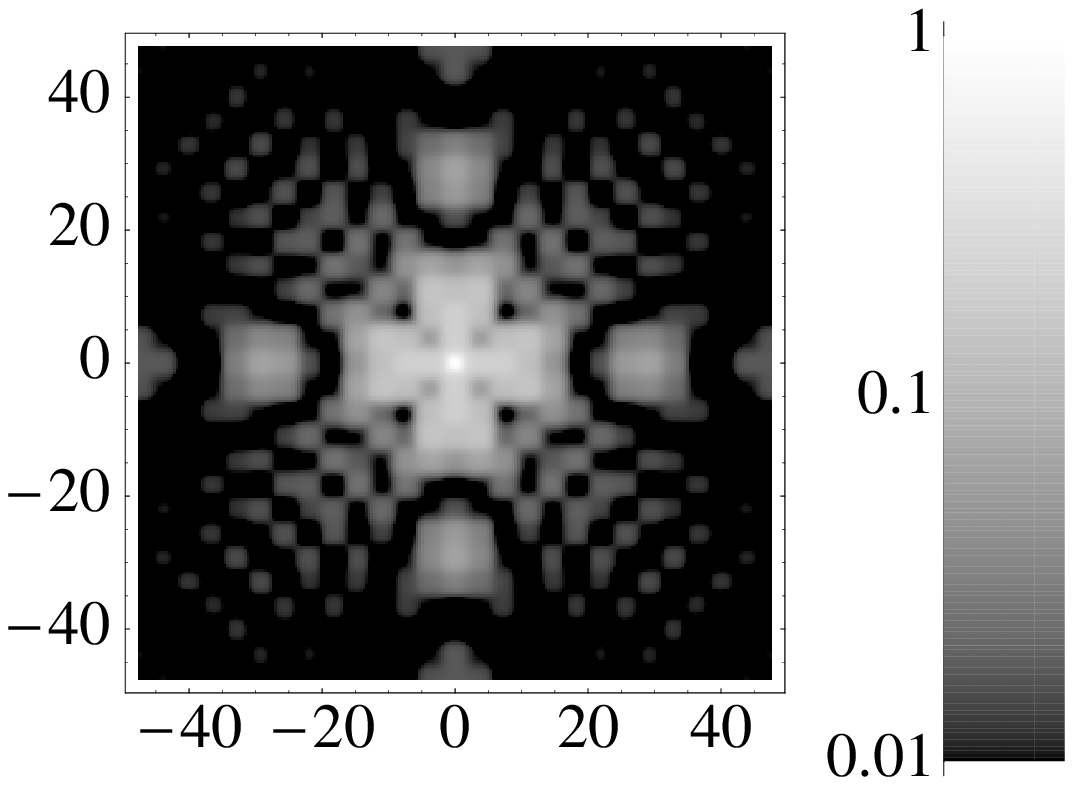}}
\caption[]{Spatial structure of the $dI/dV$ at the impurity resonance
voltage ($-1.5$~mV) for {\it iv}. Horizontal is parallel to
the gap maxima directions.
\label{planeslices}}
\end{figure}

Figure~\ref{planeslices} shows the $dI/dV$ of the resonance
for {\it iv}; {\it iii} is similar.
The differences between the Figs.~2-4 and the measurements
of Ref.~\cite{Pan}
may be due to errors in the homogeneous
electronic structure of BSCCO used in the 
calculation, particularly the low-energy electronic structure which 
dominates the longer-range LDOS. These errors may be due to inaccuracies in 
the model for the electronic structure measured 
by photoemission\cite{BSCnormal}, or they may be due 
to the neglect of other
collective effects on the surface. 
Another likely source of error is that the electronic
structure model of the host is not spin dependent.

One of the possible mechanisms of a pseudogap is local 
AF order, such as occurs in a stripe.\cite{stripe} The 
magnetic component apparent in 
the impurity potential suggests this origin as well. 
If two nearby local moments are aligned parallel, then the
resonances associated with them 
will hybridize and split\cite{twoimp}, whereas if
they are antiparallel the resonances will be degenerate. A
careful examination of the $dI/dV$ for two Zn atoms near each other
on the surface may clarify whether there is local AF order. 

The LDOS reported in Refs.~\cite{Hudson,YazdaniBSC,Pan} are
best explained by the presence of a pseudogap state on the
surface of BSCCO. The relative height of the electron-like
and hole-like resonances in the DOS depends directly on
the amplitude of local Andre\'ev processes, and thus shows
the degree of local superconducting coherence. This is of
great practical interest, for the presence of
these processes is essential to forming a proper Josephson
junction across an interface.
The information obtained about the superconducting state at
the surface of BSCCO indicates the clear promise of future
STM measurements near defects in other correlated electron
systems.

I would like to thank J. C. Davis for 
discussions and for providing the data of Ref.~\cite{Pan}. 
This work has been supported in part 
by ONR through contract No. N00014-99-1-0313.

\begin{references}
\bibitem{OPmax} Measured gap features at the surface vary
from sample to sample as well as from point to point on the
same sample. $40$~meV is a typical location for the gap
feature.
\end{references}
\end{document}